\documentclass[conference,9pt]{IEEEtran}
\IEEEoverridecommandlockouts

\usepackage{cite}
\usepackage{amsmath,amssymb,amsfonts}
\usepackage{algorithmic}
\usepackage{graphicx}
\usepackage{textcomp}
\usepackage{xcolor}
\def\BibTeX{{\rm B\kern-.05em{\sc i\kern-.025em b}\kern-.08em
    T\kern-.1667em\lower.7ex\hbox{E}\kern-.125emX}}

\usepackage{hyperref}       
\hypersetup{
    colorlinks,
    linkcolor={blue},
    citecolor={green!90!black},
    urlcolor={blue}
}

\usepackage{url}            
\usepackage{booktabs}       
\usepackage{amsfonts}       
\usepackage{xcolor}         
\usepackage{amssymb}
\usepackage{makecell}

\newcommand{\ie}{\textit{i}.\textit{e}.}
\newcommand{\eg}{\textit{e}.\textit{g}.}

\usepackage{pifont}

\usepackage{arydshln}
\usepackage{wrapfig}
\usepackage{multirow}
\usepackage[usestackEOL]{stackengine}
    
\begin{document}

\title{Contextual Speech Extraction: Leveraging Textual History as an Implicit Cue for Target Speech Extraction}

\author{\IEEEauthorblockN{Minsu Kim$^{*,1}$, Rodrigo Mira$^{*,2}$, Honglie Chen$^{1}$, Stavros Petridis$^{1,2}$, Maja Pantic$^{1,2}$\thanks{$^*$Equal contribution.}}
\vspace{0.2cm}
\IEEEauthorblockA{$^1$Meta AI, UK \quad $^2$Imperial College London, UK}
}

\maketitle
\begin{abstract}
In this paper, we investigate a novel approach for Target Speech Extraction (TSE), which relies solely on textual context to extract the target speech. We refer to this task as Contextual Speech Extraction (CSE). Unlike traditional TSE methods that rely on pre-recorded enrollment utterances, video of the target speaker's face, spatial information, or other explicit cues to identify the target stream, our proposed method requires only a few turns of previous dialogue (or monologue) history. This approach is naturally feasible in mobile messaging environments where voice recordings are typically preceded by textual dialogue that can be leveraged implicitly. We present three CSE models and analyze their performances on three datasets. Through our experiments, we demonstrate that even when the model relies purely on dialogue history, it can achieve over 90\,\% accuracy in identifying the correct target stream with only two previous dialogue turns. Furthermore, we show that by leveraging both textual context and enrollment utterances as cues during training, we further enhance our model's flexibility and effectiveness, allowing us to use either cue during inference, or combine both for improved performance. Samples and code available on \url{https://miraodasilva.github.io/cse-project-page}.
\end{abstract}


\hypersetup{linkcolor=black,urlcolor=black} 

\section{Introduction} \label{sec:intro}
Speech separation~\cite{erdogan2015phase,wang2018supervised,takahashi2020improving,li2020listen,wang2023tf} aims to separate individual speech streams from a mixture of multiple utterances in a cocktail-party scenario~\cite{cherry1953some}. Target speech extraction can be seen as a narrower version of speech separation, aimed at extracting only the target speech from a mixture, discarding all other streams. However, to identify the target stream, an external cue is necessary. In past works, this cue has taken the form of: an enrollment utterance, \ie, a pre-recorded speech sample from the target speaker~\cite{delcroix2018single,ge2022spex,zmolikova2023neural}; spatial information, \eg, the position of the target speaker with respect to the microphone, which can be estimated from a video or audio recording~\cite{DBLP:journals/taslp/GannotVGO17}; video of the target speaker's lip movements~\cite{DBLP:conf/interspeech/AfourasCZ18,mira2023lavoce,chen2024rtlavoce}; an explicit textual cue, such as a partial transcription of the target utterance~\cite{rahimi2022reading} or a prompt specifying which stream should be extracted (\eg, ``Extract the loudest speaker'')~\cite{hao2023typing,9688052,tzinis2022heterogeneous}; an image (\eg, providing an image of a boat to extract the speech related to sailing)~\cite{ohishi2022conceptbeam}; or a combination of multiple of these cues~\cite{DBLP:journals/jstsp/GuZXCZY20,hao2023typing,tzinis2022heterogeneous}. 
Notably, all of these cues are explicit, since they require the user to specify which stream should be enhanced, \eg, by acquiring a speech sample featuring only that specific speaker or by pointing the camera only at the target speaker. In practice, this places a severe constraint on the speech extraction task, as additional effort is needed to obtain this explicit cue. This raises the question: can we find an implicit cue that can enable TSE without this additional effort? To answer this, we return to the cocktail-party problem~\cite{cherry1953some} and observe that semantic information, along with popular cues such as voice samples and video~\cite{hong2022visual,pingchuan2023autoavsr,hong2023watch}, often plays a significant role in extracting relevant speech: If we know the context of the conversation, we can focus only on the speech which pertains to this specific context while suppressing all unrelated streams. Still, obtaining contextual cues in a traditional TSE scenario is typically not viable without prior knowledge, as we only have access to a mixture that contains multiple streams related to different topics, and have no way to determine which one should be extracted. Remarkably, one application where this context is readily available is mobile messaging, in which voice recordings are often interleaved with text messages, making the text-based dialogue history a naturally accessible cue to enhance these recordings. Based on this observation, we set out to build a speech extraction model that can take advantage of this textual cue.

\begin{figure}[t]
\centering
\centerline{\includegraphics[width=8.5cm]{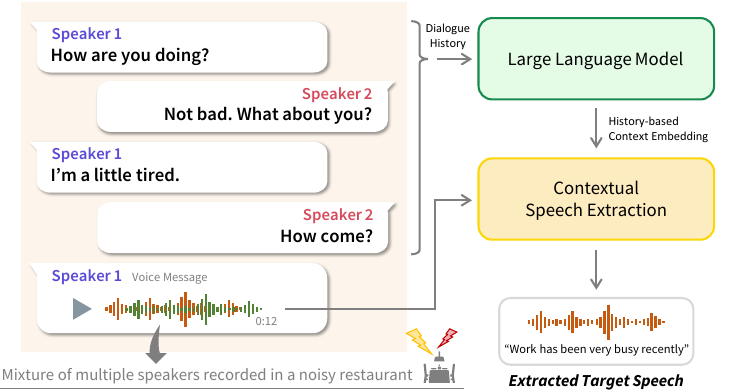}}
\vspace{-0.2cm}
\caption{Overall concept of the proposed contextual speech extraction framework: By employing dialogue history as a cue, the model can extract the target speech from a mixture without explicit cues.}
\label{fig:1}
\vspace{-0.2cm}
\end{figure}

In this paper, we introduce Contextual Speech Extraction (CSE), a novel speech extraction approach which leverages textual context as an implicit cue to infer the target speech stream from a mixture without need for any additional information about the target speaker. To investigate the feasibility of CSE, we begin by constructing a cascaded system which first separates the mixture into separate streams via a speech separation model, then transcribes each stream using an automated speech recognition (ASR) model, and finally selects the most likely stream by feeding in the context and transcription into a Large Language Model (LLM)~\cite{touvron2023llama}. Then, aiming to devise a simpler and more efficient inference procedure, we introduce a unified CSE model based on Sepformer~\cite{subakan2021sepformer} that receives the audio mixture and embedded context (from the LLM) and predicts the target speech directly. We present two variants of this model: a separator (ContSep), that separates all streams in the mixture and classifies the target stream simultaneously via the same backbone, and an extractor (ContExt), which predicts only the target stream. Furthermore, by jointly employing the enrollment utterance and the dialogue history, we build H-ContExt, a hybrid speech extraction model that can be conditioned on the enrollment utterance, the textual context, or both, yielding increased flexibility and performance. 

Our contributions can be summarized as follows:
1) To the best of our knowledge, we present the first work exploring TSE by solely employing textual history as a cue, and coin this task as CSE; 2) We investigate three CSE approaches: a cascaded method, a speech separation model, and a speech extraction model; 3) Through broad experiments conducted on three different datasets, we demonstrate the effectiveness of the proposed method for both dialogues and monologues, and analyze how different context lengths can influence speech extraction performance; 4) We show that by combining the proposed context cue with the traditional enrollment utterance, we can further enhance the performance and flexibility of our model.

\section{Contextual Speech Extraction}
Fig.~\ref{fig:1} shows the proposed CSE framework. Let $x\in\mathbb{R}^{T}$ be a speech mixture and $y\in\mathbb{R}^{T}$ be the ground-truth target speech matched with the text-based context $c$ (\eg, ``Speaker 1: How is your (...), Speaker 2: (...)''), where $T$ represents the length of the waveform. The main objective of our proposed task is to extract the target speech $y$ from the input speech mixture $x$ by conditioning on the context $c$. 

\subsection{Sepformer}
As our backbone speech separation architecture, we employ Sepformer~\cite{subakan2021sepformer}, a popular waveform-based model. It comprises an encoder, a masking network, and a decoder, with the speech separation primarily taking place in the masking network. The masking network predicts the target mask by processing the input features, which are computed from raw audio via the encoder and segmented into small chunks. Its core module is a dual-path transformer~\cite{luo2020dual,chen2020dual}, consisting of an IntraTransformer and an InterTransformer. Their roles are to encode information within a chunk (\ie, local info) and across chunks (\ie, global info), respectively, which can be put as:
\begin{equation}
\setlength{\abovedisplayskip}{3pt}
\setlength{\belowdisplayskip}{3pt}
    h_{intra} = \text{IntraTrans}(x_{intra}) \in \mathbb{R}^{N \times C \times D},
\label{eq:1}
\end{equation}
\begin{equation}
\setlength{\abovedisplayskip}{3pt}
\setlength{\belowdisplayskip}{3pt}
    h_{inter} = \text{InterTrans}(x_{inter}) \in \mathbb{R}^{C \times N \times D},
\label{eq:2}
\end{equation}
where IntraTrans($\cdot$) and InterTrans($\cdot$) represent the IntraTransformer and InterTransformer, $x_{intra}$ and $x_{inter}$ are the inputs of each transformer layer, $C$ is the chunk size, $N$ is the number of chunks, and $D$ is the dimension of embedding. The first dimension is treated as the batch in computations, meaning that short-term dependencies are modeled in Eq.~\ref{eq:1} and long-term dependencies in Eq.~\ref{eq:2}.

\begin{figure}[t]
\centering
\centerline{\includegraphics[width=1.0\linewidth]{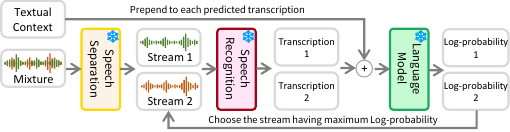}}
\vspace{-0.2cm}
\caption{Illustration of the cascaded CSE pipeline.}
\label{fig:2}
\vspace{-0.2cm}
\end{figure}

\subsection{Cascaded CSE} \label{sec:cascaded}
We start with a simple approach by building a cascaded model that connects three different models, as depicted in Fig.~\ref{fig:2}: a pre-trained Sepformer~\cite{subakan2021sepformer} for speech separation, Whisper-base~\cite{radford2023whisper} for speech recognition, and Llama 3-8B~\cite{touvron2023llama} for language modeling. These modules are combined via a four-step procedure: 1) the input mixed speech is separated into multiple candidate streams via the Sepformer; 2) The transcription of each separated stream is obtained using Whisper; 3) The conversation history $c$ is prepended to each transcription and passed into Llama 3-8B, which computes the log-probability of each utterance; 4) The utterance showing the highest log-probability is chosen as the predicted stream. 

\subsection{Unified CSE}
Although it serves its purpose as a naive approach to CSE, the cascaded model's reliance on three separate pre-trained models incurs salient redundancies (\eg, transcribing all separated streams, which is computationally costly) which decrease the efficiency of the inference procedure and can lead to error propagation. To mitigate these issues, we merge the cascade by training a unified model that can directly predict the target speech while implicitly modeling the textual context as a condition. We modify the architecture of the Sepformer to incorporate contextual information when performing TSE. First, we obtain the context embedding $f_c$ by passing the text of the dialogue/monologue history $c$ into an LLM (Llama3-8B~\cite{touvron2023llama}) and extracting the network's last hidden state. We then incorporate $f_c$ into the masking network by concatenating it with the input of each IntraTransformer and InterTransformer layer along the temporal dimension. This can be expressed as:
\begin{equation}
\setlength{\abovedisplayskip}{3pt}
\setlength{\belowdisplayskip}{3pt}
    h_{intra} = \text{IntraTrans}(f_c \oplus x_{intra}) \in \mathbb{R}^{N \times (C+1) \times D},
\label{eq:3}
\end{equation}
\begin{equation}
\setlength{\abovedisplayskip}{3pt}
\setlength{\belowdisplayskip}{3pt}
    h_{inter} = \text{InterTrans}(f_c \oplus x_{inter}) \in \mathbb{R}^{C \times (N+1) \times D},
\label{eq:4}
\end{equation}
where $\oplus$ represents concatenation along the temporal dimension. The additional temporal dimension induced by the concatenation is detached before passing it to the next transformer layer, so that the context embedding can be injected directly into the next layer as well. We present two variants of this unified model: \textbf{Cont}extual \textbf{Sep}arator (ContSep) and \textbf{Cont}extual \textbf{Ext}ractor (ContExt).

\begin{figure}[t]
\centering
\centerline{\includegraphics[width=8.5cm]{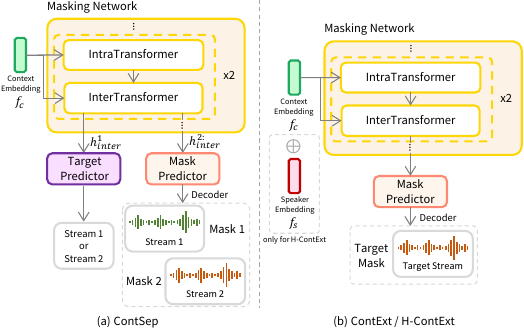}}
\vspace{-0.2cm}
\caption{Illustration of the proposed CSE methods. (a) ContSep that separates all streams from the mixture and predicts the target stream, (b) ContExt and H-ContExt that produce the target stream only. For simplicity, the encoder and decoder are omitted in the figure.}
\label{fig:3}
\vspace{-0.2cm}
\end{figure}

\textbf{ContSep} is shown in Fig.~\ref{fig:3}a. It separates individual speech streams from the input mixture, similar to a standard speech separator, while also including a target predictor that predicts which stream is most likely the target stream based on the context. In this way, we can identify the target speech while also obtaining the non-target streams, which can be useful in scenarios where we would also like to extract non-target speech. In practice, the model learns to perform separation and classification simultaneously by minimizing $\mathcal{L}$, which is the sum of the cross-entropy loss and the permutation-invariant negative Scale-Invariant Signal-to-Noise Ratio (SI-SNR) loss~\cite{kolbaek2017multitalker,roux2019sisnr,luo2019conv}:
\begin{equation}
\setlength{\abovedisplayskip}{3pt}
\setlength{\belowdisplayskip}{3pt}
    \mathcal{L} = - \log p(\hat{\mathbf{y}}_t) + \min_{\sigma \in \mathfrak{S}_n} \sum_i^n \mathcal{L}_{\text{SI-SNR}}(\mathbf{y}_i, \mathbf{\hat{y}}_{\sigma(i)}),
\end{equation}
\begin{equation}
\setlength{\abovedisplayskip}{3pt}
\setlength{\belowdisplayskip}{3pt}
    \mathcal{L}_{\text{SI-SNR}}(y, \hat{y}) = -10\log_{10}\frac{\|\bar{y}\|^2}{\|\varepsilon\|^2},
\end{equation}
\begin{equation}
\setlength{\abovedisplayskip}{3pt}
\setlength{\belowdisplayskip}{3pt}
    \bar{y} = \frac{{<}\hat{y},y{>}y}{\|y\|^2}, \quad \varepsilon=\hat{y}-\bar{y},
\end{equation}

where $p(\hat{\mathbf{y}}_t)$ is the probability of selecting the index of the target stream $t$ by the target predictor, $\mathbf{y}=y_{1 \dots n}$ and $\mathbf{\hat{y}}=\hat{y}_{1 \dots n}$ represent the $n$ ground-truth source speeches of the mixure and the $n$ predicted streams respectively, and ${\sigma}{\in}{\mathfrak{S}_n}$ indicates the permutation indexes of $n$ elements. In short, the permutation loss calculates the negative SI-SNR loss, $\mathcal{L}_{\text{SI-SNR}}$ for all possible pairs of ground-truth and predicted speech streams and choose the pairs showing the minimum loss. The first sequence of hidden states from the last layer of the masking network $h_{inter}^1$ is employed to predict the target stream by averaging out the chunk dimension, while the remaining hidden states $h_{inter}^{2:N+1}$ are used to predict the mask. As we employ the permutation invariant loss, there is no way to know which output stream is the target stream before generating the output speeches. To resolve this ambiguity, we measure the SI-SNR between the predicted waveforms $\mathbf{\hat{y}}$ and the target stream $y$, and choose the waveform with the highest SI-SNR as our target label $t$.

\textbf{ContExt} is shown in Fig.~\ref{fig:3}b.  Although ContSep can provide both predicted target stream and non-target streams, its performance may be bottlenecked by having to separate all speech streams from the mixture, compared to typical TSE models that focus solely on producing the target stream. Therefore, we design a simple yet effective model which focuses solely on the target stream, aligning its objectives with those of traditional TSE models~\cite{delcroix2020improving,ji2020speaker}. Unlike ContSep, ContExt only contains one mask predictor which predicts the mask for the target stream only and there is no need to consider permutation. Therefore, its learning process is simpler, minimizing only the negative SI-SNR loss, $\mathcal{L}_{\text{SI-SNR}}$ measured between the target speech $y$ and the predicted speech $\hat{y}$.

\subsection{Hybrid CSE with Joint Vocal and Contextual Cues}
In an attempt to alleviate the enrollment utterance constraint of traditional TSE models and leverage our findings above, we investigate a powerful and flexible TSE method by jointly modeling the textual context and enrollment utterance as cues. We refer to this model as \textbf{H}ybrid \textbf{Cont}extual \textbf{Ext}ractor (H-ContExt), and depict it in Fig.~\ref{fig:3}b. Concretely, we extract the speaker embedding $f_s$ from the enrollment utterance by employing a pre-trained speaker verification model~\cite{desplanques2020ecapa,ravanelli2021speechbrain}, and concatenate it with context embedding $f_c$ as:
\begin{equation}
\setlength{\abovedisplayskip}{3pt}
\setlength{\belowdisplayskip}{3pt}
    h_{intra} = \text{IntraTrans}(f_c \oplus f_s \oplus x_{intra}) \in \mathbb{R}^{N \times (C+2) \times D},
\end{equation}
\begin{equation}
\setlength{\abovedisplayskip}{3pt}
\setlength{\belowdisplayskip}{3pt}
    h_{inter} = \text{InterTrans}(f_c \oplus f_s \oplus x_{inter}) \in \mathbb{R}^{C \times (N+2) \times D},
\end{equation}
where the embedding dimensions of $f_c$ and $f_s$ are mapped to $D$ by using a linear layer for each. During training, we randomly mask out one of the cues (\ie, $f_c$ or $f_s$), to prevent the model from relying only on the speaker embedding, and enabling inference using either the textual context, enrollment utterance, or both.

\section{Experiments}
\subsection{Datasets and Metrics}
We train and evaluate the proposed method on three transcribed speech datasets (two for dialogue and one for monologue) with continuous long-form recordings, to ensure long textual context\,\footnote{Only non-Meta authors conducted any of the dataset preprocessing (no dataset pre-processing took place on Meta’s servers or facilities).}.

\textbf{DailyTalk}~\cite{lee2023dailytalk} is a spoken dialogue dataset whose dialogues are derived from DailyDialog~\cite{li2017dailydialog}. It contains 2,\,541 dialogues uttered by two speakers (one male and one female), totaling 20 hours of speech. The data is randomly split into 2,\,285 dialogues for training, and 128 dialogues each for validation and testing.

\textbf{SpokenWOZ}~\cite{si2024spokenwoz} is a task-oriented spoken dialogue dataset containing 5,\,700 dialogues, uttered by 250 speakers, totaling 249 hours. We adopt the train/val/test split provided by the authors. 

\textbf{TED-LIUM 3}~\cite{hernandez2018ted} a spoken monologue dataset originally proposed for ASR containing 2,\,351 recording sessions collected from TED talks with an average duration of 11m\,30s. The aim of our TED-LIUM 3 experiments is to evaluate whether the proposed CSE can be applied in a single-speaker lecture-style scenario, as an alternative to dialogue-based messaging environments. We use the `speaker adaptation' split proposed by the authors, ensuring that the speakers in the training set do not overlap with those in the evaluation set.

During evaluation on SpokenWoz and TED-LIUM, to ensure sufficient context length, we include only samples that have at least 10 utterances of past context. For DailyTalk, given its shorter dialogues, the minimum number of past utterances is set to 5 instead. 

\textbf{Metrics.} To measure the speech separation/extraction performance, we use two established metrics: SI-SNR improvement (SI-SNR\,i)~\cite{roux2019sisnr} and Signal-to-Distortion Ratio improvement (SDR\,i)~\cite{vincent2006performance}. In addition, we report the accuracy (ACC) of the model in selecting the correct target stream. Accuracy is measured by calculating the SI-SNR between the extracted speech and all speech sources in the mixture (\ie, ground-truth), and choosing the highest as the model's prediction.

\begin{table*}[t]
\renewcommand{\arraystretch}{1.2}
\renewcommand{\tabcolsep}{2.0mm}
\caption{Speech extraction performance on 2-speaker mixtures from DailyTalk, SpokenWOZ, and TED-LIUM 3.}
\vspace{-0.2cm}
\centering
\resizebox{0.95\linewidth}{!}{
\begin{tabular}{l c ccc ccc ccc}
\Xhline{3\arrayrulewidth}
\multirow{2}{*}{\textbf{Method}} & \multirow{2}{*}{\textbf{Cue}} & \multicolumn{3}{c}{\textbf{DailyTalk}} & \multicolumn{3}{c}{\textbf{SpokenWOZ}} & \multicolumn{3}{c}{\textbf{TED-LIUM 3}}\\ \cmidrule(l{2pt}r{2pt}){3-5} \cmidrule(l{2pt}r{2pt}){6-8} \cmidrule(l{2pt}r{2pt}){9-11}
 & & \textbf{SI-SNR\,i (dB)} & \textbf{SDR\,i (dB)} & \textbf{ACC (\%)} & \textbf{SI-SNR\,i (dB)} & \textbf{SDR\,i (dB)} & \textbf{ACC (\%)}& \textbf{SI-SNR\,i (dB)} & \textbf{SDR\,i (dB)} & \textbf{ACC (\%)} \\ \cmidrule(l{2pt}r{2pt}){1-2} \cmidrule(l{2pt}r{2pt}){3-5} \cmidrule(l{2pt}r{2pt}){6-8} \cmidrule(l{2pt}r{2pt}){9-11}
\multicolumn{11}{l}{$\bullet$ \textbf{\textit{Speech Separation}}} \\
\,\,\, Sepformer \cite{subakan2021sepformer} & Oracle & 19.66 & 20.13 & 100 & 14.15 & 14.60 & 100 & 16.06 & 16.29 & 100 \\ 
\,\,\, Sepformer \cite{subakan2021sepformer} & Random & -11.56 & 1.34 & 50.0 & -9.41 & -1.47 & 50.0 & -10.41 & -2.47 & 50.0 \\ \hdashline

\multicolumn{11}{l}{$\bullet$ \textbf{\textit{Target Speech Extraction (TSE)}}} \\
\,\,\, Sepformer~\cite{desplanques2020ecapa} & Audio & 19.98 & 20.51 & 99.8 & 15.99 & 16.61 & 97.9 & 14.38 & 14.92 & 96.4 \\ \hdashline

\multicolumn{11}{l}{$\bullet$ \textbf{\textit{Contextual Speech Extraction (CSE) - Ours}}} \\
\,\,\, Cascaded & Context & 0.74 & 8.51 & 70.5 & -1.65 & 3.86 & 66.2 & 5.52 & \textbf{8.96} & 79.7 \\
\,\,\, ContSep & Context & 5.42 & \textbf{11.55} & 76.9 & 13.97 & 14.78 & 97.3 & 5.60 & 8.66 & \textbf{79.9} \\
\,\,\, ContExt & Context & \textbf{10.03} & 11.35 & \textbf{83.5} & \textbf{14.52} & \textbf{15.01} & \textbf{98.3} & \textbf{5.71} & 6.46 & 76.6 \\ 
\hdashline
\multicolumn{11}{l}{$\bullet$ \textbf{\textit{Hybrid Contextual Speech Extraction (TSE\,+\,CSE) - Ours}}} \\
\,\,\, H-ContExt & Context & 10.81 & 12.18 & 80.9 & 14.93 & 15.50 & 96.2 & 4.66 & 6.48 & 74.2 \\
\,\,\, H-ContExt & Audio & 19.95 & 20.48 & 99.8 & 15.47 & 16.14 & 97.5 & 15.26 & 15.63 & 97.8 \\
\,\,\, H-ContExt & Both (A+C) & \textbf{20.00} & \textbf{20.54} & \textbf{99.8} & \textbf{15.98} & \textbf{16.53} & \textbf{98.5} & \textbf{15.37} & \textbf{15.73} & \textbf{97.9} \\
\Xhline{3\arrayrulewidth}
\end{tabular}}
\label{table:1}
\vspace{-0.1cm}
\end{table*}

\subsection{Implementation Details}
We follow the original Sepformer architecture~\cite{subakan2021sepformer}, where 8 layers with embedding dimension $D{=}256$ are employed for each IntraTransformer and InterTransformer, with 2 Intra-Inter blocks in total. The encoder and decoder comprise a single 1D convolution layer each, and the chunk size ($C$) is set to 250 with 50\,\% overlap. The input speech is resampled to 8\,kHz for all experiments. During training, we mix the target and interfering streams with a random SNR between -5 and 5\,dB~\cite{zeghidour2021wavesplit}, and apply random speed perturbation with ratios of 0.9, 1.0, and 1.1, random acoustic noise perturbation from the DEMAND dataset~\cite{thiemann2013diverse} with 0--10 dB SNR, and random time shifting with a maximum shift of 1s. For DailyTalk, we train all unified CSE models for 300\,k steps with learning rate warmup (5\,k steps) and decay~\cite{loshchilov2022sgdr}, setting the peak learning rate to $1.5\times10^{-4}$ and the batch size to 16. For SpokenWOZ and TED-LIUM 3, all models are trained similarly for 500\,k steps with 10\,k warmup steps, a batch size of 32, and a peak learning rate of $3\times10^{-4}$. Unified CSE models are initialized from our Sepformer trained on the same dataset. We use AdamW~\cite{kingma2015adam,loshchilovdecoupled} for optimization.

\subsection{Results}
The performances of our TSE models on the three benchmark databases are shown in Table~\ref{table:1}. 
For comparison purposes, we evaluate the Sepformer's~\cite{subakan2021sepformer} speech extraction performance by either assuming we know which stream is the target stream (Oracle) or randomly picking it from separated streams (Random). To provide a strong TSE baseline, we also train a TSE equivalent of ContExt by feeding the speaker embedding~\cite{desplanques2020ecapa}, extracted from an enrollment utterance, into our model instead of the context embedding.

\textbf{Comparison between CSE methods.} 
By observing the performance of our CSE models, we confirm that they all achieve adequate speech extraction accuracy by employing only the context cue. Among the three, ContExt shows the best performance on DailyTalk and SpokenWOZ. As expected, the cascaded model suffers from error propagation, particularly evident in its very poor results on SpokenWOZ. Upon further analysis, we find that the main source of errors is the ASR model: Since SpokenWOZ features recordings by non-native speakers, the ASR model struggles to produce accurate transcriptions, resulting in poor performance. To quantify this, we measure the Word Error Rate (WER) of the ASR model on the ground-truth test sets of each dataset, which are 9.99\,\%, 38.59\,\%, and 8.90\,\% for DailyTalk, SpokenWOZ, and TED-LIUM, respectively. Alternatively, by developing unified CSE models (\ie, ContSep and ContExt) that forego the need for speech recognition, we can achieve consistent results regardless of the speakers' accents. On TED-LIUM, where native speakers are dominant, we can see that the cascaded approach is effective (although highly redundant, as discussed above), achieving slightly better SDR\,i than its peers, while ContSep attains the highest accuracy, and ContExt maintains the highest SI-SNR\,i. 

\begin{figure}[t]
\centering
\centerline{\includegraphics[width=1.0\linewidth]{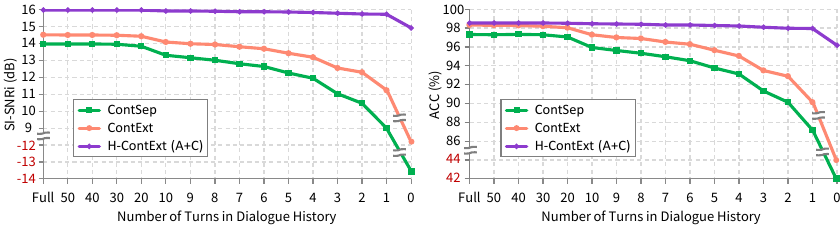}}
\vspace{-0.1cm}
\caption{Performances (Left: SI-SNR\,i, Right: ACC) of Unified CSE and Hybrid CSE on different context lengths on SpokenWOZ dataset. ``Full'' indicates that we use the full context, without limiting its length.}
\label{fig:4}
\vspace{-0.2cm}
\end{figure}

\textbf{Hybrid CSE.} 
The proposed H-ContExt offers the flexibility to perform TSE using context, audio, or both cues, making it adaptable for real-world applications. Moreover, by employing both cues, H-ContExt achieves better performance compared to audio-only TSE methods, highlighting its effectiveness.

\textbf{Ablation on context length.}
To analyze the performance of the proposed CSE models with different context lengths, we measure their performance by limiting the number of turns in the dialogue history in Fig.~\ref{fig:4}. Each turn of SpokenWOZ dataset contains an average of 8.7 words. We can see that the performance rises sharply with the inclusion of more utterances when the number of turns is small, and then begins to plateau around context length 20. Remarkably, by employing only two previous dialogue turns, we achieve more than 90\,\% ACC in choosing the right stream without employing other cues, demonstrating the viability of CSE with short context lengths. Finally, when no dialogue history is given, the model cannot perform properly, which is expected since no cue is provided.

\begin{table}[t]
	\renewcommand{\arraystretch}{1.2}
	\renewcommand{\tabcolsep}{2.2mm}
\caption{Performance on 3-speaker mixtures from TED-LIUM 3.}
\vspace{-0.2cm}
\centering
\resizebox{0.95\linewidth}{!}{
\begin{tabular}{l c ccc}
\Xhline{3\arrayrulewidth}
\textbf{Method} & \textbf{Cue} & \textbf{SI-SNR\,i (dB)} & \textbf{SDR\,i (dB)} & \textbf{ACC (\%)} \\ \cmidrule(l{2pt}r{2pt}){1-2} \cmidrule(l{2pt}r{2pt}){3-5}
\multicolumn{5}{l}{$\bullet$ \textbf{\textit{Speech Separation}}} \\
\,\,\, Sepformer \cite{subakan2021sepformer} & Oracle & 14.12 & 14.41 & 100 \\
\,\,\, Sepformer \cite{subakan2021sepformer} & Random & -11.36 & -5.67 & 33.3 \\\hdashline

\multicolumn{5}{l}{$\bullet$ \textbf{\textit{Target Speech Extraction (TSE)}}} \\
\,\,\, Sepformer \cite{desplanques2020ecapa} & Audio & 5.02 & 8.21 & 71.1 \\ \hdashline

\multicolumn{5}{l}{$\bullet$ \textbf{\textit{Contextual Speech Extraction (CSE) - Ours}}} \\
\,\,\, Cascaded & Context & -1.62 & 2.20 & 59.3 \\
\,\,\, ContSep & Context & -0.87 & 3.54 & 62.2 \\
\,\,\, ContExt & Context & \textbf{2.61} & \textbf{4.70} & \textbf{62.8} \\ 
\hdashline

\multicolumn{5}{l}{$\bullet$ \textbf{\textit{Hybrid Speech Extraction (TSE + CSE) - Ours}}} \\
\,\,\, H-ContExt & Context & 3.41 & 5.30 & 64.0 \\
\,\,\, H-ContExt & Audio & 4.62 & 7.26 & 70.7 \\
\,\,\, H-ContExt & Both (A+C) & \textbf{5.18} & \textbf{7.61} & \textbf{71.6} \\
\Xhline{3\arrayrulewidth}
\end{tabular}}
\label{table:2}
\vspace{-0.1cm}
\end{table}
\begin{table}[t]
	\renewcommand{\arraystretch}{1.2}
	\renewcommand{\tabcolsep}{3mm}
\caption{Ablation study using different sizes of LLM to encode the dialogue history-based context embedding on DailyTalk.}
\vspace{-0.2cm}
\centering
\resizebox{0.8\linewidth}{!}{
\begin{tabular}{c ccc}
\Xhline{3\arrayrulewidth}
\textbf{Method} & \textbf{SI-SNR\,i (dB)} & \textbf{SDR\,i (dB)} & \textbf{ACC (\%)} \\ \cmidrule(l{2pt}r{2pt}){1-1} \cmidrule(l{2pt}r{2pt}){2-4}
OPT-350M \cite{zhang2022opt} & 6.52 & 8.19 & 76.02 \\
OPT-1.3B \cite{zhang2022opt} & 8.25 & 10.02 & 79.01 \\
OPT-2.7B \cite{zhang2022opt} & 8.24 & 9.59 & 79.66 \\
Llama 3-8B \cite{touvron2023llama} & \textbf{10.03} & \textbf{11.35} & \textbf{83.51} \\
\Xhline{3\arrayrulewidth}
\end{tabular}}
\label{table:3}
\vspace{-0.3cm}
\end{table}

\textbf{3-Speaker Mixtures.}
To validate the robustness of our CSE models, we evaluate the performance of the proposed CSE on 3-speaker mixtures from TED-LIUM 3 in Table \ref{table:2}. Our findings confirm that our models can extract the target speech using only the context cue for 3-speaker cases as well, achieving over 60\,\% ACC (compared to 33.3\,\% ACC for random selection). Unlike in the 2-speaker scenario, ContExt convincingly outperforms its CSE counterparts for 3-speaker mixtures from TED-LIUM 3. This is likely due to ContSep and the cascaded approach having to separate all streams rather than simply extracting the target speech, which means their performance degrades more noticeably with the addition of more background speakers.

\textbf{Ablation on different LLMs.}
To evaluate how different Large Language Models (LLMs) can affect CSE performance, we experiment with a range of LLMs of different sizes to extract the context embedding. Specifically, we experimented with OPT-350M, OPT-1.3B, OPT-2.7B, and Llama~3-8B~\cite{zhang2022opt,touvron2023llama} to train ContExt on the DailyTalk dataset. The ablation results are shown in Table \ref{table:3}. We see that, for the same model type (OPT), employing larger LLMs consistently yields better results, and conclude that our largest LLM, Llama 3-8B, achieves the best results overall.

\section{Conclusion}
We propose Contextual Speech Extraction (CSE), a new task that relies on textual context to perform target speech extraction, and design three models for this purpose. Through our experiments on three different datasets (two for dialogue and one for monologue), we demonstrate the effectiveness of CSE without employing explicit cues. Furthermore, by combining our newfound contextual cue with the established enrollment utterance, we present a flexible and effective new type of hybrid TSE model that can leverage textual context and/or enrollment utterances to extract target speech from a mixture.

\clearpage
\bibliographystyle{IEEEtran}
\bibliography{egbib}

\end{document}